\documentclass[11pt]{article}
\usepackage{amsmath}
\usepackage{amsfonts}
\usepackage{amsthm}
\usepackage{graphicx}

\newtheorem{theorem}{Theorem}[section]
\newtheorem{proposition}[theorem]{Proposition}
\newtheorem{corollary}[theorem]{Corollary}
\newtheorem{lemma}[theorem]{Lemma}
\newenvironment{definition}{\medskip\noindent{\it Definition:\/} }{\medskip}
\newenvironment{remark}{\medskip\noindent{\it Remark:\/} }{\medskip}
\numberwithin{equation}{section}

\newcommand{\diag}{{\rm diag}}
\newcommand{\Tr}{{\rm Tr}}
\begin{document}
\title{Random matrices with external source and multiple
orthogonal polynomials}
\author{P.M. Bleher and A.B.J. Kuijlaars}
\date{July 25, 2003}
\maketitle

\begin{abstract}
We show that the average characteristic polynomial
$P_n(z) = \mathbb E [\det(zI-M)]$ of the
random Hermitian matrix ensemble
$Z_n^{-1} \exp(-\Tr(V(M)-AM))dM$ is characterized by
multiple orthogonality conditions that depend on the eigenvalues of
the external source $A$. For each eigenvalue $a_j$ of $A$,
there is a weight and $P_n$ has $n_j$ orthogonality conditions
with respect to this weight, if $n_j$ is the multiplicity of $a_j$.
The eigenvalue correlation functions have determinantal form,
as shown by Zinn-Justin. Here we give a different expression for
the kernel. We derive a Christoffel-Darboux formula in case $A$ has
two distinct eigenvalues, which leads to a compact formula
in terms of a Riemann-Hilbert problem that is satisfied by
multiple orthogonal polynomials.
\end{abstract}

\section{Random matrices with external source}

Following Br\'ezin and Hikami \cite{BH1,BH2,BH3,BH4} and
P.~Zinn-Justin \cite{ZJ1,ZJ2} we consider a random matrix ensemble
with an external source,
\begin{equation} \label{prob2}
    \frac{1}{Z_n} e^{-\Tr(V(M) - AM)} dM
\end{equation}
defined on $n\times n$ Hermitian matrices $M$.
The ensemble (\ref{prob2}) consists of a general unitary invariant
part $V(M)$ and an extra term $AM$ where $A$ is a fixed
$n \times n$ Hermitian matrix, the external source or the external
field. Due to the external source, the ensemble (\ref{prob2}) is not
unitary invariant.
For the special Gaussian case $V(x) = \frac{1}{2} x^2$, we
can write $M$ in (\ref{prob2}) as
$M = H + A$
where $H$ is a random matrix from the GUE ensemble, and $A$
is deterministic, hence in this case it reduces to the class
of deterministic plus random matrices studied in
\cite{Pas,BHZ,BH1,BH2,BH3,BH5,BH4}.

Zinn-Justin \cite{ZJ1} showed that the eigenvalue correlations
of ensemble (\ref{prob2}) can be expressed in the determinantal form,
\[ R_m(\lambda_1, \ldots, \lambda_m)
    = \det(K_n(\lambda_i,\lambda_j))_{i,j=1,\ldots,m} \]
for some kernel $K_n$. In this paper, we give a different expression
for $K_n$. We believe that our formulation is useful for asymptotic
analysis. Indeed, for the Gaussian case $V(x) = \frac{1}{2}x^2$ and
for the case where $A$ has only two distinct eigenvalues, we have
been able to carry out the asymptotic analysis almost completely.
This will be reported elsewhere.

Our approach is based on the observation  that the average
characteristic polynomial
\[ P_n(z) = \mathbb E \left[ \det(z-M) \right] \]
of the ensemble (\ref{prob2}) can be characterized by
the property that
\[ \int_{-\infty}^{\infty} P_n(x) x^k e^{-(V(x) - a_jx)} dx = 0 \]
for every eigenvalue $a_j$ of $A$ and for $k = 0, \ldots, n_j-1$,
where $n_j$ is the multiplicity of $a_j$, see Section 2.
We can embed the polynomial $P_n$ in a sequence of polynomials
$\{ P_k \}_0^n$ where $P_k$ has degree $k$. Then our kernel has the form
\begin{equation} \label{kernelKn}
    K_n(x,y) = e^{-\frac{1}{2} (V(x) + V(y))}
    \sum_{k=0}^{n-1} P_k(x) Q_k(y)
\end{equation}
where the $Q_k$ are certain dual functions (not polynomials in general),
see Section 3.

When $A=0$ (no external source), the polynomials $P_k$
are usual monic orthogonal polynomials with respect to the
weight $e^{-V(x)}$ on $\mathbb R$. In that case, the
function $Q_k$ is a multiple of $P_k$ and the kernel (\ref{kernelKn})
reduces to the orthogonal polynomial kernel which is familiar
in the theory of random matrices. By the Christoffel-Darboux
formula we then have
\begin{equation} \label{CDformula0}
    K_n(x,y) = e^{-\frac{1}{2}(V(x)+V(y))} \frac{\gamma_{n-1}}{\gamma_{n}}
    \frac{P_n(x) P_{n-1}(y) - P_{n-1}(x) P_n(y)}{x-y},
\end{equation}
where $\gamma_k$ is the leading coefficient of the
orthonormal polynomial of degree $k$.

In Section 4 we present an analog of the Christoffel-Darboux
formula for the kernel (\ref{kernelKn}) in the case where $A$
has only two eigenvalues.
We also relate it to a Riemann-Hilbert problem in Section 5.

\section{The average characteristic polynomial}
We define the monic polynomial
\[ P_n(z) = \mathbb E \left[ \det(z-M) \right] \]
where the expectation is with respect to the ensemble (\ref{prob2}).

\begin{proposition} \label{prop21}
Suppose $A$ has eigenvalues $a_j$, $j=1,\ldots,n$ with
$a_i \neq a_j$ if $i \neq j$.
Then the following hold.
\begin{enumerate}
\item[\rm (a)] There is a constant $\tilde{Z}_n$ such that
\begin{equation} \label{formulaPn}
P_n(z) = \frac{1}{\tilde{Z}_n}
    \int \prod_{j=1}^n (z-\lambda_j)
    \prod_{j=1}^n e^{-(V(\lambda_j) - a_j \lambda_j)} \Delta(\lambda) d\lambda,
\end{equation}
where
\[ \Delta(\lambda) = \prod_{i>j} (\lambda_i-\lambda_j) \]
and $d \lambda = d\lambda_1 d\lambda_2 \cdots d\lambda_n$.
\item[\rm (b)] Let
\[ m_{jk} = \int_{-\infty}^{\infty} x^k e^{-(V(x) - a_j x)} d x. \]
Then we have the determinantal formula
\begin{equation}
    \label{detformulaPn}
    P_n(z) = \frac{1}{\tilde{Z}_n}
    \begin{vmatrix}
    m_{10} & m_{11} & \cdots & m_{1n} \\
    \vdots & \vdots & \ddots & \vdots \\
    m_{n0} & m_{n1} & \cdots & m_{nn} \\
    1 & z &  \cdots &  z^n
    \end{vmatrix}.
\end{equation}
\item[\rm (c)] For $j=1, \ldots, n$,
\begin{equation} \label{multortrelations}
    \int_{-\infty}^{\infty} P_n(x) e^{-(V(x)-a_jx)} dx = 0,
    \end{equation}
and these equations uniquely determine the monic polynomial $P_n$.
\end{enumerate}
\end{proposition}

\begin{proof}
Write $M = U \Lambda U^*$ where $U$ is unitary and $\Lambda$ is
diagonal, $\Lambda = \diag(\lambda_1, \ldots, \lambda_n)$.
of $M$ on the diagonal.
Then by the Weyl integration formula, see e.g.\ \cite{Deift,Mehta},
we have for every integrable function $f$ on the space of Hermitian
$n\times n$ matrices,
\begin{equation} \label{Weyl}
    \int f(M) dM = \pi^{-n(n-1)/2} \left(\prod_{j=0}^n j!\right)
        \iint f(U\Lambda U^*)  \Delta(\lambda)^2 d\lambda dU
\end{equation}
where $dU$ denotes the normalized Haar measure on the unitary group $U(n)$.
Thus
\[ P_n(z) = \frac{\left(\prod_{j=0}^n j!\right) }{Z_n \pi^{n(n-1)/2}}  \int \prod_{j=1}^n (z-\lambda_j)
    \prod_{j=1}^n e^{-V(\lambda_j)}
    \left(\int e^{AU \Lambda U^*} dU \right) \Delta(\lambda)^2 d\lambda. \]
Because of  the Harish-Chandra, Itzykson--Zuber integral \cite{HC,IZ}
\[ \int e^{AU \Lambda U^*} dU = \left(\prod_{j=0}^{n-1} j! \right)
    \frac{\det\left( e^{a_j \lambda_k}\right)}{\Delta(a)\Delta(\lambda)}, \]
we obtain that
\begin{equation} \label{formintPn}
    P_n(z) = \frac{n!\left(\prod_{j=0}^{n-1} j!\right)^2}{Z_n \pi^{n(n-1)/2}}
    \int \prod_{j=1}^n (z-\lambda_j)
    \prod_{j=1}^n e^{-V(\lambda_j)}
    \frac{\det\left(e^{a_j \lambda_k}\right)}{\Delta(a)} \Delta(\lambda) d\lambda.
\end{equation}
We expand the determinant
\[ \det\left(e^{a_j \lambda_k}\right) = \sum_{\sigma \in S_n}
    (-1)^{\sigma}
    \prod_{j=1}^n e^{a_j \lambda_{\sigma(j)}} \]
where $S_n$ is the symmetric group.  Hence
\[ P_n(z) =\frac{n!\left(\prod_{j=0}^{n-1} j!\right)^2}{Z_n \pi^{n(n-1)/2} \Delta(a)}
    \sum_{\sigma \in S_n} (-1)^{\sigma}
    \int \prod_{j=1}^n (z-\lambda_j)
    \prod_{j=1}^n e^{-V(\lambda_j)}
    \prod_{j=1}^n e^{a_j \lambda_{\sigma(j)}}
    \Delta(\lambda) d\lambda. \]
We make a change of variables
$\lambda'_j = \lambda_{\sigma(j)} $. Then
$(-1)^{\sigma} \Delta(\lambda) = \Delta(\lambda')$, hence in the sum over
$S_n$, we have $n!$ equal terms and, by dropping the prime, we obtain
(\ref{formulaPn}) with constant
\begin{equation} \label{tildeZn}
    \tilde{Z}_n = Z_n \frac{ \pi^{n(n-1)/2} \Delta(a)}{\left(\prod_{j=0}^{n} j!\right)^2}.
\end{equation}
 This proves part (a).
\medskip

Observe that
\[ \prod_{j=1}^n (z-\lambda_j) \Delta(\lambda) =
    \Delta(\lambda,z) =
    \begin{vmatrix}
    1 & \lambda_1 & \cdots & \lambda_1^n \\
    \vdots & \vdots & \ddots & \vdots \\
    1 & \lambda_n & \cdots & \lambda_n^n \\
    1 & z & \cdots & z^n
    \end{vmatrix} \]
and
\[ \prod_{j=1}^n (z-\lambda_j) \prod_{j=1}^n e^{-(V(\lambda_j) - a_j\lambda_j)}
    \Delta(\lambda) =
    \begin{vmatrix}
    e^{-(V(\lambda_1) - a_1\lambda_1)} & \lambda_1e^{-(V(\lambda_1) - a_1\lambda_1)}
    & \cdots & \lambda_1^n e^{-(V(\lambda_1) - a_1\lambda_1)} \\
    \vdots & \vdots & \ddots & \vdots \\
    e^{-(V(\lambda_n) - a_n\lambda_n)} & \lambda_ne^{-(V(\lambda_n) - a_n\lambda_n)}
    & \cdots & \lambda_n^n e^{-(V(\lambda_n) - a_n\lambda_n)} \\
    1 & z & \cdots & z^n
    \end{vmatrix}.
\]
Then (\ref{detformulaPn}) follows immediately from this and (\ref{formulaPn}).
This proves part (b).
\medskip

From (\ref{detformulaPn}) it follows that
\[ \int_{-\infty}^{\infty} P_n(x) e^{-(V(x)-a_jx)} dx
    = \frac{1}{\tilde{Z}_n}
    \begin{vmatrix}
    m_{10} & m_{11} & \cdots & m_{1n} \\
    \vdots & \vdots & \ddots & \vdots \\
    m_{n0} & m_{n1} & \cdots & m_{nn} \\
    m_{j0} & m_{j1} & \cdots & m_{jn}
    \end{vmatrix} = 0, \]
for every $j=1, \ldots, n$. This proves (\ref{multortrelations}).
To prove uniqueness of $P_n$ satisfying (\ref{multortrelations}),
observe that by equating the coefficients of $x^n$ in (\ref{detformulaPn})
we obtain that
\begin{equation} \label{detformulatildeZn}
    \begin{vmatrix}
    m_{10} & m_{11} & \cdots & m_{1,n-1} \\
    \vdots & \vdots & \ddots & \vdots \\
    m_{n0} & m_{n1} & \cdots & m_{n,n-1}
    \end{vmatrix} = \tilde{Z}_n \neq 0.
\end{equation}
Let $P_n(x) = x^n + p_{n-1} x^{n-1} + \cdots + p_0$ and set
$p = \begin{pmatrix} p_0 & \cdots & p_{n-1} \end{pmatrix}^T$.
Then the equations (\ref{multortrelations}) are written in terms
of the vector $p$ as
\[ Mp= -m, \qquad M = \left(m_{jk}\right)_{j=1,\ldots,n;k=0,\ldots,n-1},
    \qquad m = (m_{jn})_{j=1,\ldots,n}. \]
By (\ref{detformulatildeZn}), $\det M \neq 0$, hence $p$ and therefore $P_n$ is unique.
\end{proof}

Proposition \ref{prop21} can be extended to the case of
multiple $a_j$s as follows.

\begin{proposition} \label{prop22}
Suppose $A$ has distinct eigenvalues $a_i$, $i=1,\ldots,p$ with
respective multiplicities $n_i$ so that $n_1 + \cdots + n_p = n$.
Let $n^{(i)} = n_1 + \cdots + n_i$ and $n^{(0)} = 0$. Define
\[ w_j(x) = x^{d_j-1} e^{-(V(x)-a_ix)}, \qquad j=1,\ldots, n, \]
where $i = i_j$ is such that $n^{(i-1)} < j \leq n^{(i)}$ and
$d_j = j-n^{(i-1)}$. Then the following hold.
\begin{enumerate}
\item[\rm (a)] There is a constant $\tilde{Z}_n > 0$ such that
\begin{equation} \label{formulaPn2}
P_n(z) = \frac{1}{\tilde{Z}_n}
    \int \prod_{j=1}^n (z-\lambda_j)
    \prod_{j=1}^n w_j(\lambda_j)  \Delta(\lambda) d\lambda.
\end{equation}
\item[\rm (b)] Let
\[ m_{jk} = \int_{-\infty}^{\infty} x^k w_j(x) dx. \]
Then we have the determinantal formula
\begin{equation}
    \label{detformulaPn2}
    P_n(z) = \frac{1}{\tilde{Z}_n}
    \begin{vmatrix}
    m_{10} & m_{11} & \cdots & m_{1n} \\
    \vdots & \vdots & \ddots & \vdots \\
    m_{n0} & m_{n1} & \cdots & m_{nn} \\
    1 & z &  \cdots &  z^n
    \end{vmatrix}.
\end{equation}
\item[\rm (c)] For $i=1, \ldots, p$,
\begin{equation} \label{multortrelations2}
    \int_{-\infty}^{\infty} P_n(x) x^j e^{-(V(x)-a_ix)} dx = 0,
    \qquad j=0, \ldots, n_i-1,
    \end{equation}
and these equations uniquely determine the monic polynomial $P_n$.
\end{enumerate}
\end{proposition}

\begin{proof}
We write
\[ \alpha = (\alpha_1, \alpha_2, \ldots, \alpha_n) =
     ( \underbrace{a_1, \ldots, a_1}_{n_1 \textrm{ times}}, a_2, \ldots,
    a_{p-1},
    \underbrace{a_p, \ldots, a_p}_{n_p \textrm{ times}}) \]
Apply formula (\ref{formintPn}) in the case when all $a_j = \tilde{a}_j$
are different and take a limit to the multiple $a_j$'s. In this limit
we have that
\[ \lim \frac{\det \left(e^{\tilde{a}_j \lambda_k}\right)}{\Delta(\tilde{a})}
    = \frac{\det \left(\lambda_k^{d_j-1} e^{\alpha_j \lambda_k}\right)}
    {\Delta_0(a) \prod_{i=1}^p \prod_{k=1}^{n_i-1} k!} \]
where $d_j$ is as in the statement of the proposition,
and
\[ \Delta_0(a) = \prod_{i > j} (a_i - a_j)^{n_in_j}. \]
Thus, formula (\ref{formintPn}) becomes
\begin{equation} \label{formintPn2}
    P_n(z) = \frac{n!\left(\prod_{j=0}^{n-1} j!\right)^2}{Z_n \pi^{n(n-1)/2}}
    \int \prod_{j=1}^n (z-\lambda_j)
    \prod_{j=1}^n e^{-V(\lambda_j)}
    \frac{\det\left(\lambda_k^{d_j-1} e^{\alpha_j \lambda_k}\right)}
    {\Delta_0(a) \prod_{i=1}^p \prod_{k=1}^{n_i-1} k!} \Delta(\lambda) d\lambda.
\end{equation}
Then we continue as in the proof of Proposition \ref{prop21}, that is,
we write
\[ \det\left(\lambda_k^{d_j-1} e^{\alpha_j \lambda_k}\right) =
    \sum_{\sigma \in S_n} (-1)^{\sigma} \prod_{j=1}^n
    \lambda_{\sigma(j)}^{d_j-1} e^{\alpha_j \lambda_{\sigma(j)}}, \]
and insert this into (\ref{formintPn2}) to obtain a sum of $n!$ equal terms,
which leads to (\ref{formulaPn2}) with
\begin{equation}
    \tilde{Z}_n = Z_n
    \frac{\pi^{n(n-1)/2}
    \Delta_0(a) \prod_{i=1}^p\prod_{k=1}^{n_i-1} k!}
    {\left(\prod_{j=0}^{n} j!\right)^2}.
\end{equation}
This proves part (a).
\medskip

Parts (b) and (c) follow from (\ref{formulaPn2}) in the same way as parts
(b) and (c) of Proposition \ref{prop21} followed from (\ref{formulaPn}).
Note that in particular we  have as in
(\ref{detformulatildeZn}),
\begin{equation} \label{detformulatildeZn2}
    \tilde{Z}_n =
    \begin{vmatrix}
    m_{10} & m_{11} & \cdots & m_{1,n-1} \\
    \vdots & \vdots & \ddots & \vdots \\
    m_{n0} & m_{n1} & \cdots & m_{n,n-1}
    \end{vmatrix} \neq 0.
\end{equation}
\end{proof}

\begin{remark} Formula (\ref{formulaPn2}) can be also written
in the following form:
\begin{equation} \label{formulaPn2a}
P_n(z) = \frac{1}{\hat{Z}_n}
    \int \prod_{j=1}^n (z-\lambda_j)
    \prod_{j=1}^n e^{-(V(\lambda_j)-a_{i_j}\lambda_j)}
    \prod_{i=1}^p \Delta(\lambda^{(i)})\,  \Delta(\lambda) d\lambda,
\end{equation}
where
$\lambda^{(i)}
=\left(\lambda_{n^{(i-1)}+1},\dots,\lambda_{n^{(i)}}\right)$
and
\begin{equation} \label{formulaPn2b}
\hat{Z}_n=\tilde{Z}_n n_1!\dots n_p!.
\end{equation}
When $A=0$, (\ref{formulaPn2a})
 reduces to the usual formula for $P_n(z)$
with respect to the  random matrix ensemble without external source.
\end{remark}

\begin{corollary}
Under the same assumptions as in Proposition \ref{prop22}, we have that
\begin{equation} \label{nextmoment}
    \int_{-\infty}^{\infty} P_n(x) x^{n_i} e^{-(V(x)-a_ix)} dx \neq 0
\end{equation}
for $i =1,\ldots, p$.
\end{corollary}
\begin{proof}
Let $P_{n+1}$ be the average characteristic polynomial
of an ensemble of $(n+1) \times (n+1)$ Hermitian random matrices whose
external source has the same eigenvalues
as $A$ plus an additional eigenvalue $a_i$. Then by part (c) of
Proposition \ref{prop22} we have that $P_{n+1}$ is the unique monic
polynomial that satisfies
the relations (\ref{multortrelations}) with $n_i$ replaced by $n_i+1$.
If $\int_{-\infty}^{\infty} P_n(x) x^{n_i} e^{-(V(x)-a_ix)} dx$ would
vanish, then $P_{n+1} + P_n$ would satisfy these relations as well, which
would contradict the uniqueness of $P_{n+1}$.
\end{proof}

\begin{remark}
The relations (\ref{multortrelations}) can be viewed as
multiple orthogonality conditions for the polynomial $P_n$.
There are $p$ weights $e^{-(V(x)-a_jx)}$, $j=1,\ldots, p$,
and for each weight there are a number of orthogonality
conditions, so that the total number of them is $n$.
This point of view is especially useful in case $A$
has only a small number of distinct eigenvalues. We will come
back to this in Section 5 when we are considering the case of
two distinct eigenvalues in detail.

There is a considerable literature on multiple orthogonal
polynomials (also called Hermite-Pad\'e polynomials), see
e.g.\ \cite{Aptekarev,ABVA,VAGK} and the references therein.
\end{remark}

\section{Determinantal form of joint probability density function}

As in Proposition \ref{prop22}, we assume that $A$ is a fixed
Hermitian matrix whose eigenvalues
$a_1, \ldots, a_p$ have respective multiplicities $n_1, \ldots, n_p$,
so that $\sum_{i=1}^p n_i = n$.
We let $\Sigma_n$ be the collection of functions
\begin{equation} \label{Sigman}
    \Sigma_n := \{ x^{j} e^{a_i x} \mid i=1, \ldots, p, \, j=0, \ldots, n_i-1 \}.
\end{equation}
We start with a lemma.
\begin{lemma} \label{lemma31}
There exists a unique function $Q_{n-1}$ in the linear span of $\Sigma_n$ such
that
\begin{equation} \label{charQn1}
    \int_{-\infty}^{\infty} x^j Q_{n-1}(x) e^{-V(x)} dx = 0
        \qquad \mbox{ for } j =0, \ldots, n-2,
\end{equation}
and
\begin{equation} \label{charQn2}
    \int_{-\infty}^{\infty} x^{n-1} Q_{n-1}(x) e^{-V(x)}dx = 1.
\end{equation}
\end{lemma}
\begin{proof}
The conditions (\ref{charQn1}) and (\ref{charQn2}) give us
$n$ linear equations for the $n$ coefficients of $Q_{n-1}$
with respect to the basis $\Sigma_n$ with coefficient matrix
\[ \begin{pmatrix}
    m_{10} & m_{20} & \ldots & m_{n0} \\
    m_{11} & m_{21} & \ldots & m_{n1} \\
    \vdots & \vdots & \ddots & \vdots \\
    m_{1,n-1} & m_{2,n-1} & \ldots & m_{n,n-1}
    \end{pmatrix} \]
where $m_{jk}$ is as in part (b) of Proposition \ref{prop22}. This
matrix is non-singular by (\ref{detformulatildeZn2}), so that the
linear equations have a unique solution, and therefore $Q_{n-1}$ exists
and  is unique.
\end{proof}

For the rest of this section, we choose some ordering of the
eigenvalues of $A$ taking into account the multiplicities, say
\begin{equation} \label{ordering} \alpha_1, \alpha_2, \ldots, \alpha_n,
\end{equation}
so that each $a_i$ appears exactly $n_i$ times among the $\alpha$'s.
For each $k = 0, 1, \ldots, n$, we can construct $P_k$ as in the
previous section, but based on $\alpha_1, \ldots, \alpha_k$.
Thus $P_k$ is a monic polynomial of degree $k$ such that
\begin{equation} \label{defPk}
    \int_{-\infty}^{\infty} P_k(x) x^j e^{-(V(x) - a_ix)} dx = 0, \qquad i=1, \ldots, p,
\quad j = 0, \ldots, k_i-1,
\end{equation}
where $k_i$ is the number of times that $a_i$ appears among
$\alpha_1, \ldots, \alpha_k$. We also have tht $P_k$ is the average characteristic polynomial
of the ensemble of $k\times k$ Hermitian matrices with external source
having eigenvalues $a_i$ with multiplicity $k_i$.

For each $k = 1, \ldots, n$ we also have by Lemma \ref{lemma31} a function $Q_{k-1}$
from the linear span of the functions
\begin{equation} \label{Sigmak}
    \Sigma_{k} := \{ x^j e^{a_ix} \mid i=1,\ldots, p, \, j = 0, \ldots, k_i-1 \}.
\end{equation}
such that
\begin{equation} \label{charQk1}
    \int_{-\infty}^{\infty} x^i Q_{k-1}(x) e^{-V(x)} dx = 0 \qquad \mbox{ for } i =0, \ldots, k-2
\end{equation}
and
\begin{equation} \label{charQk2}
    \int_{-\infty}^{\infty} x^{k-1} Q_{k-1}(x) e^{-V(x)}dx = 1.
\end{equation}
It follows from (\ref{defPk}), (\ref{charQk1}), and (\ref{charQk2}) that
the $P$'s and $Q$'s are a biorthogonal system in the sense that
\begin{equation} \label{biorthogonal}
    \int_{-\infty}^{\infty} P_j(x) Q_k(x) e^{-V(x)} dx = \delta_{jk},
    \qquad \textrm{ for } j,k=0, \ldots, n-1.
\end{equation}
This property explains why we used $Q_{k-1}$ for the function that
satisfies (\ref{charQk1}) and (\ref{charQk2}) (and not $Q_k$).

We now introduce the kernel $K_n$.

\begin{definition}
With the polynomials $P_k$ and the functions $Q_k$ introduced above,
we define
\begin{equation} \label{defKn}
    K_n(x,y) = e^{- \frac{1}{2} (V(x)+ V(y))}
    \sum_{k=0}^{n-1} P_k(x) Q_k(y).
\end{equation}
\end{definition}
Note that the $P$'s and the $Q$'s depend on the specific ordering
(\ref{ordering}) that we choose for the eigenvalues of $A$. However,
it will turn out that $K_n$ does not depend on this ordering.
\medskip

Because of the biorthogonality property (\ref{biorthogonal}) it is easy to see
from the definition (\ref{defKn}) that we have
\begin{equation} \label{diagonal}
    \int_{-\infty}^{\infty} K_n(x,x) dx = n
\end{equation}
and the reproducing kernel property
\begin{equation} \label{repkernel}
    \int_{-\infty}^{\infty} K_n(x,y) K_n(y,z) dy = K_n(x,z).
\end{equation}

The following is the main theorem of this paper.
\begin{theorem}
The joint probability density function on eigenvalues has the
determinantal form
\begin{equation} \label{jpdf}
    \frac{1}{n!} \det (K_n(\lambda_j, \lambda_k))_{1\leq j,k\leq n}
\end{equation}

The $m$-point correlation function has the form
\begin{equation} \label{Rm}
    R_m(\lambda_1, \ldots, \lambda_m) =
    \det (K_n(\lambda_j,\lambda_k))_{1\leq j,k\leq m}
\end{equation}
\end{theorem}

\begin{proof}
Any joint probability density function of the form (\ref{jpdf})
with a kernel $K_n$ satisfying (\ref{diagonal}) and
(\ref{repkernel}) leads to $m$-point correlation functions
of the form (\ref{Rm}). So it suffices to prove
that (\ref{jpdf}) is the joint probability density function
of the eigenvalues.

For each $j$, we define
\begin{equation} \label{defwj}
    w_j(x) = x^{d_j-1} e^{a_ix}
\end{equation}
if $a_i = \alpha_j$ and $a_i$ appears $d_j$ times
in the sequence $\alpha_1, \ldots, \alpha_j$. Note that the
functions (\ref{defwj}) differ from the functions $w_j$ used
in Proposition \ref{prop22} in two respects. First there is an
extra factor $e^{-V(x)}$ in Proposition \ref{prop22}, and second we
used a specific ordering of the eigenvalues of $A$ in Proposition
\ref{prop22} (which only amounts to a renumbering).

A similar calculation as that leading to (\ref{formintPn}) in the proof of
Proposition \ref{prop21} shows
that the joint probability density of eigenvalues is proportional to
\[ \prod_{j=1}^n e^{-V(\lambda_j)}
    \det (w_i(\lambda_j))_{1\leq i,j \leq n}
    \Delta(\lambda). \]
Since $Q_{i-1}$ is a linear combination of $w_1, \ldots, w_i$ we
can take appropriate row combinations to find that
\[ \det(w_i(\lambda_j))_{1\leq i,j\leq n} \propto \det (Q_{i-1}(\lambda_j))_{1\leq i,j \leq n}. \]
We write $\Delta(\lambda)$ as a Vandermonde determinant
which we similarly rewrite as
\[ \Delta(\lambda)
    = \det (P_{i-1}(\lambda_k))_{1\leq i,k\leq n}. \]
Thus the joint probability density of eigenvalues is proportional to
\[
    \det \left(e^{-\frac{1}{2} V(\lambda_j)} Q_{i-1}(\lambda_j)\right)_{1\leq i,j\leq n}
    \det \left(e^{-\frac{1}{2} V(\lambda_k)} P_{i-1}(\lambda_k)\right)_{1\leq i,k\leq n}.
\]
Taking the transpose of the matrix in the first determinant, and then using the
multiplicative property of determinants, we find that the joint probability
density is equal to
\[ c\det(K_n(\lambda_j,\lambda_k))_{1\leq j,k\leq n} \]
for some constant $c$, which should be such that the integral
with respect to $d\lambda_1 \cdots d\lambda_n$
is $1$. Because of the properties (\ref{diagonal}) and (\ref{repkernel})
this is so for $c = \frac{1}{n!}$ and the theorem is proved.
\end{proof}

\begin{remark}
Renumbering the eigenvalues $a_1, a_2, \ldots a_n$
leads to the same kernel $K_n$ but to different $P_k$ and
$Q_k$.
\end{remark}

\section{Special form of the kernel in case of two eigenvalues}

In this section we assume we have only two distinct eigenvalues
$a_1$ and $a_2$ with multiplicities $n_1$ and $n_2$, respectively.
We order the eigenvalues $\alpha_1, \alpha_2, \ldots, \alpha_n$
in some arbitrary way ($a_j$ appear $n_j$ times in the sequence),
but for convenience we assume that
\begin{equation} \label{assumption1}
    \alpha_{n-1} = a_1 \qquad \alpha_n = a_2.
\end{equation}
We also put
\begin{equation} \label{assumption2}
    \alpha_{n+1} = a_1 \qquad \alpha_{n+2} = a_2.
\end{equation}
As in the preceding section, we have polynomials $P_k$ and
functions $Q_k$ for every $k=0, \ldots, n-1$ such that
\[ K_n(x,y) = e^{-\frac{1}{2} (V(x)+ V(y))}
    \sum_{k=0}^{n-1} P_k(x) Q_k(y). \]
It is our aim in this section to simplify this expression.
The formula we will find is an analogue of the well-known Christoffel-Darboux
formula for orthogonal polynomials.

To present the formula we are going to use multi-index notation.
For non-negative integers $k_1$ and $k_2$, we use $P_{k_1,k_2}$ to
denote the monic polynomial of degree $k_1+k_2$ having
$k_j$ orthogonality relations with respect to the weight
\[ w_j(x) = e^{-(V(x)-a_jx)}, \qquad j=1,2. \]
Thus
\begin{equation} \label{orthogonalitytypeII}
    \int_{-\infty}^{\infty} P_{k_1,k_2}(x) x^i w_j(x) dx = 0,
    \qquad i=0, \ldots,k_j-1, \ j=1,2.
\end{equation}
The polynomial $P_{k_1,k_2}$ is called a multiple orthogonal polynomial
of type II, see e.g.\ \cite{Aptekarev,ABVA}.
We also define
\begin{equation} \label{typeI}
    Q_{k_1,k_2}(x) = A_{k_1,k_2}(x) e^{a_1x} + B_{k_1,k_2}(x) e^{a_2x}
\end{equation}
where the degree of $A_{k_1,k_2}$ is $k_1-1$, the degree of $B_{k_1,k_2}$ is $k_2-1$
and
\begin{equation} \label{orthogonalitytypeI}
    \int_{-\infty}^{\infty} x^j Q_{k_1,k_2}(x) e^{-V(x)} dx =
        \left\{ \begin{array}{cl} 0, &  j=0, \ldots, k_1+k_2-2,\\[10pt]
            1 & j= k_1+k_2-1. \end{array} \right.
\end{equation}
The polynomials $A_{k_1,k_2}$ and $B_{k_1,k_2}$ are called multiple
orthgonal polynomials of type I, see \cite{Aptekarev,ABVA}.
For each pair $(k_1,k_2)$ of non-negative integers, the polynomials
$P_{k_1,k_2}$, $A_{k_1,k_2}$, and $B_{k_1,k_2}$ exist and are uniquely
defined by their degree requirements and the relations
(\ref{orthogonalitytypeII}), (\ref{typeI}), and (\ref{orthogonalitytypeI}).

We can express $P_k$ and $Q_k$ in this new notation as
\[ P_k = P_{k_1,k_2}, \qquad \textrm{and} \qquad Q_{k-1} = Q_{k_1,k_2} \]
provided $a_j$ appears $k_j$ times among the numbers
$\alpha_1, \ldots, \alpha_k$ (for $j=1,2$).
In particular, we have because of our assumptions (\ref{assumption1}) and
(\ref{assumption2})
\begin{equation} \label{formulasP}
    P_n = P_{n_1,n_2}, \quad P_{n-1} = P_{n_1,n_2-1}, \quad P_{n-2} = P_{n_1-1,n_2-1}.
\end{equation}
and
\begin{equation} \label{formulasQ}
    Q_{n-1} = Q_{n_1,n_2}, \quad Q_n = Q_{n_1+1,n_2}, \quad Q_{n+1} = Q_{n_1+1, n_2+1}.
\end{equation}

We also need the numbers
\begin{equation} \label{h-numbers}
    h_{k_1,k_2}^{(j)} =
    \int_{-\infty}^{\infty} P_{k_1,k_2}(x) x^{k_j} w_j(x) dx, \qquad j=1,2,
\end{equation}
which are non-zero, cf.\ (\ref{nextmoment}).
For later use we note that
\begin{eqnarray*}
    1 & = & \int P_{k_1,k_2}(x) Q_{k_1+1,k_2}(x) e^{-V(x)} dx \\
    & = &  \int P_{k_1,k_2}(x) (A_{k_1+1,k_2}(x) w_1(x) + B_{k_1+1,k_2}(x) w_2(x)) dx \\
    & = & \int P_{k_1,k_2}(x) A_{k_1+1,k_2}(x) w_1(x) dx \\
    & = & \left(\textrm{leading coefficient of } A_{k_1+1,k_2} \right) \times h_{k_1,k_2}^{(1)}
\end{eqnarray*}
so that
\begin{equation} \label{leadingA}
    \textrm{leading coefficient of } A_{k_1+1,k_2} = \frac{1}{h_{k_1,k_2}^{(1)}}.
\end{equation}
Similarly,
\begin{equation} \label{leadingB}
    \textrm{leading coefficient of } B_{k_1,k_2+1} = \frac{1}{h_{k_1,k_2}^{(2)}}.
\end{equation}
It also follows from (\ref{leadingA}) and (\ref{leadingB}) that
$h_{k_1,k_2}^{(j)} \neq 0$ for $j=1,2$.

Then we can state the following theorem.
\begin{theorem} \label{theorem41}
With the notation introduced above, we have
\begin{eqnarray} \nonumber
    (x-y) e^{\frac{1}{2} (V(x) + V(y))} K_n(x,y) & = &
      P_{n_1,n_2}(x) Q_{n_1,n_2}(y)   \\
     && \nonumber  - \frac{h_{n_1,n_2}^{(1)}}{h_{n_1-1,n_2}^{(1)}} P_{n_1-1,n_2}(x) Q_{n_1+1,n_2}(y) \\
     && \label{CDformula}  - \frac{h_{n_1,n_2}^{(2)}}{h_{n_1,n_2-1}^{(2)}} P_{n_1,n_2-1}(x) Q_{n_1,n_2+1}(y)
\end{eqnarray}
\end{theorem}

The proof of the theorem needs some preparation.
We start working again with the $P_k$'s and $Q_j$'s (single index) as
before.
For each $j$ and $k$, we put
\[ c_{jk} = \int x P_k(x) Q_j(x) e^{-V(x)} dx. \]
The coefficients $c_{jk}$ appear in the expansion
\begin{equation} \label{xPk}
    xP_k(x) = \sum_{j=0}^{k+1} c_{jk} P_j(x),
\end{equation}
since by the biorthogonality relation we have indeed
\[ c_{jk} = \int xP_k(x) Q_j(x) e^{-V(x)} dx. \]
Similarly, we have for $j=0, \ldots, n-1$,
\begin{equation} \label{xQj}
    xQ_j(x) = \sum_{k=0}^{n+1} c_{jk}  Q_k(x).
\end{equation}
Note that by adding the two values $\alpha_{n+1}$ and $\alpha_{n+2}$
as we did in (\ref{assumption2}), we have this expansion for every $j \leq n-1$.

\begin{lemma} \label{lemma42}
\begin{enumerate}
\item[\rm (a)] If $j \geq k+2$ then $c_{jk} = 0$.
\item[\rm (b)] If $k \geq j+3$ and if both $a_1$ and
$a_2$ appear at least once among
$\alpha_{j+2}, \alpha_{j+3}, \ldots, \alpha_k$, then $c_{jk} = 0$.
\end{enumerate}
\end{lemma}
\begin{proof}
(a) We have that
\[ \int P(x) Q_j(x) e^{-V(x)} dx = 0 \]
for every polynomial $P$ of degree $\leq j-1$.
Since $xP_k$ is a polynomial of degree $k+1$, it follows
that $c_{jk}=0$ if $k+1 \leq j-1$. This proves part (a).

(b) Let $k$ and $j$ be such that the conditions of part (b)
are satisfied. Suppose that $a_1$  appears $k_1$ times
among $\alpha_1, \ldots, \alpha_k$, and $j_1$ times among
$\alpha_1, \ldots, \alpha_{j+1}$. We put $k_2 = k-k_1$ and $j_2 = j+1-j_1$.
It follows from the assumptions that $j_1 < k_1$ and $j_2 < k_2$.
Then
$Q_{j}(x) = Q_{j_1,j_2}(x) = A_{j_1,j_2}(x) e^{a_1x} + B_{j_1,j_2}(x) e^{a_2x}$
where $A_{j_1,j_2}$ has degree $j_1-1$ and $B_{j_1,j_2}$ has degree $j_2-1$.
It follows that
\[ xQ_j(x) = xA_{j_1,j_2}(x) e^{a_1x} + x B_{j_1,j_2}(x) e^{a_2x} \]
and $xA_{j_1,j_2}(x)$ has degree $j_1 \leq k_1-1$ and $x B_{j_1,j_2}(x)$ has
degree $j_2 \leq k_2-1$. Thus, by the multiple orthogonality
property of $P_k = P_{k_1,k_2}$, we have
\[ \int P_k(x) x Q_j(x) e^{-V(x)} dx = 0. \]
This proves part (b).
\end{proof}

We also need the following relations between near-by $P$'s and $Q$'s.
\begin{lemma} \label{lemma43}
We have
\begin{eqnarray}
    P_{n_1-1,n_2-1} & = & \nonumber
    \frac{h_{n_1-1,n_2-1}^{(1)}}{h_{n_1-1,n_2}^{(1)}}
    \left(P_{n_1-1,n_2} - P_{n_1,n_2-1}\right) \\
    & = & \label{relationP}
    - \frac{h_{n_1-1,n_2-1}^{(2)}}{h_{n_1,n_2-1}^{(2)}}
    \left(P_{n_1-1,n_2} - P_{n_1,n_2-1}\right)
\end{eqnarray}
and
\begin{eqnarray}
Q_{n_1+1,n_2+1} & = & \nonumber
    - \frac{h_{n_1,n_2}^{(1)}}{h_{n_1,n_2+1}^{(1)}} \left(Q_{n_1,n_2+1} - Q_{n_1+1,n_2}\right) \\
    & = & \label{relationQ}
    \frac{h_{n_1,n_2}^{(2)}}{h_{n_1+1,n_2}^{(2)}} \left(Q_{n_1,n_2+1} - Q_{n_1+1,n_2}\right)
\end{eqnarray}
\end{lemma}
\begin{proof}
Since $P_{n_1-1,n_2}$ and $P_{n_1,n_2-1}$ are both monic polynomials
of degree $n$, their difference is a polynomial of degree $\leq n-1$.
Since this difference  has $n_j-1$ orthogonality conditions
with respect to $w_j$ for $j=1,2$, it must be a multiple of $P_{n_1-1,n_2-1}$.
Thus
\[  P_{n_1-1,n_2} - P_{n_1,n_2-1} = \gamma P_{n_1-1,n_2-1} \]
for some $\gamma$.
Integrating this equation with respect to $x^{n_1-1} w_1(x)$ and
$x^{n_2-1} w_2(x)$, we get
$h_{n_1-1,n_2}^{(1)} = \gamma h_{n_1-1,n_2-1}^{(1)}$, and
$-h_{n_1,n_2-1}^{(2)} = \gamma h_{n_1-1,n_2-1}^{(2)}$,
respectively.
This gives (\ref{relationP}).

Next we note that  we have
\[ \int x^j (Q_{n_1,n_2+1} - Q_{n_1+1,n_2}) e^{-V(x)} dx = 0  - 0 = 0,
    \qquad j=0, \ldots, n_1+n_2-1 \]
and also
\[ \int x^{n_1+n_2} (Q_{n_1,n_2+1} -Q_{n_1+1,n_2}) e^{-V(x)} dx
    = 1- 1  = 0.  \]
Since $Q_{n_1,n_2+1}(x) - Q_{n_1+1,n_2}(x) = A(x) e^{a_1x} + B(x) e^{a_2x}$
where $A$ has degree $n_1$ and $B$ has degree $n_2$, it follows that
$Q_{n_1,n_2+1} - Q_{n_1+1,n_2}$ is a multiple of $Q_{n_1+1,n_2+1}$, say
\[ Q_{n_1,n_2+1} - Q_{n_1+1,n_2} = \beta Q_{n_1+1,n_2+1}. \]
This means for  the $A$-polynomials that
\[ A_{n_1,n_2+1} - A_{n_1+1,n_2} = \beta A_{n_1+1,n_2+1} \]
and looking at the leading coefficient (= coefficient of $x^{n_1}$)
we get
\[ \beta = - \frac{\textrm{leading coefficient of }A_{n_1+1,n_2}}
    {\textrm{leading coefficient of }A_{n_1+1,n_2+1}}
    = - \frac{h_{n_1,n_2+1}^{(1)}}{h_{n_1,n_2}^{(1)}},
\]
where we used (\ref{leadingA}).
We also get by considering the $B$-polynomials that
\[ \beta = \frac{\textrm{leading coefficient of } B_{n_1,n_2+1}}
    {\textrm{leading coefficient of }B_{n_1+1,n_2+1}}
    = \frac{h_{n_1+1,n_2}^{(2)}}{h_{n_1,n_2}^{(2)}} \]
because of (\ref{leadingB}). This proves (\ref{relationQ}).
\end{proof}

Now we are ready for the proof of Theorem \ref{theorem41}.
\begin{proof}
We note that $(x-y) e^{\frac{1}{2}(V(x)+V(y))} K_n(x,y)$ has a telescoping character.
Indeed we have by (\ref{xPk}) and (\ref{xQj}),
\begin{eqnarray*}
    (x-y) \sum_{k=0}^{n-1} P_k(x) Q_k(y) & = &
    \sum_{k=0}^{n-1} xP_k(x) Q_k(y)
    - \sum_{j=0}^{n-1} y P_j(x) Q_j(y) \\
    & = &
    \sum_{k=0}^{n-1} \sum_{j=0}^{k+1} c_{jk} P_j(x) Q_k(y)
    - \sum_{j=0}^{n-1} \sum_{k=0}^{n+1} c_{jk} P_j(x) Q_k(y) \\
    & = &
    c_{n,n-1} P_n(x) Q_{n-1}(y)
    - \sum_{j=0}^{n-1} c_{jn} P_j(x) Q_n(y)
    - \sum_{j=0}^{n-1} c_{j,n+1} P_j(x) Q_{n+1}(y).
\end{eqnarray*}
Now observe that $c_{n,n-1} = 1$, and that $c_{jn} = 0$ for
$j=0, \ldots, n-3$ and $c_{j,n+1} = 0$ for $j=0, \ldots, n-2$,
which follows from Lemma \ref{lemma42}.
Thus
\begin{eqnarray} \nonumber
 (x-y) e^{\frac{1}{2} (V(x) +V(y))} K_n(x,y) & = &
    P_n(x) Q_{n-1}(y) \\
    & & \nonumber
    - c_{n-2,n} P_{n-2}(x) Q_n(y)  \\
    & & \nonumber - c_{n-1,n} P_{n-1}(x) Q_n(y) \\
    & &- c_{n-1,n+1} P_{n-1}(x) Q_{n+1}(y).
    \label{formulaKn1}
\end{eqnarray}
In formula (\ref{formulaKn1}) we have reduced the $n$-term
expression to four terms, which is already quite nice.
However, we want to reduce to three terms only.
Changing back to multi-index notation and using (\ref{formulasP}) and (\ref{formulasQ}),
we see that (\ref{formulaKn1}) leads to
\begin{eqnarray} \nonumber
(x-y) e^{\frac{1}{2}(V(x)+ V(y))} K_n(x,y)
   & = & P_{n_1,n_2}(x) Q_{n_1,n_2}(y) \\
   & & \nonumber
        - c_{n-2,n} P_{n_1-1,n_2-1}(x) Q_{n_1+1,n_2}(y) \\
        & & \nonumber
        - c_{n-1,n} P_{n_1,n_2-1}(x) Q_{n_1+1,n_2}(y) \\
        & &- c_{n-1,n+1} P_{n_1,n_2-1}(x) Q_{n_1+1,n_2+1}(y)
        \label{formulaKn2}
\end{eqnarray}

Comparing (\ref{formulaKn2}) and (\ref{CDformula}) we see that we need
to get rid of $P_{n_1-1,n_2-1}$ and $Q_{n_1+1,n_2+1}$.
This can be done using the following relations between near-by $P$'s and $Q$'s.

Our next task is to express the recurrence coefficients
$c_{n-2,n}$, $c_{n-1,n}$, and $c_{n-1,n+1}$ that appear in (\ref{formulaKn2})
in terms of the $h$-numbers. This is rather straightforward from the definition.
Indeed, we have
\begin{eqnarray} \nonumber
c_{n-2,n} & = &
\int x P_{n_1,n_2}(x) Q_{n_1,n_2-1}(x) e^{-V(x)} dx \\
& = & \nonumber
\int P_{n_1,n_2}(x) \left( x A_{n_1,n_2-1}(x) w_1(x) + x B_{n_1,n_2-1}(x) w_2(x)\right) dx \\
& = & \nonumber
\int P_{n_1,n_2}(x) x A_{n_1,n_2-1}(x) w_1(x) dx \\
& = & \nonumber
\left(\textrm{leading coefficient of } A_{n_1,n_2-1}\right) \times h_{n_1,n_2}^{(1)} \\
& = & \label{formulacn-2n}
\frac{h_{n_1,n_2}^{(1)}}{h_{n_1-1,n_2-1}^{(1)}} \end{eqnarray}
where we used (\ref{leadingA}),
and similarly,
\begin{eqnarray} \label{formulacn-1n}
c_{n-1,n} =   \frac{h_{n_1,n_2}^{(1)}}{h_{n_1-1,n_2}^{(1)}}
    + \frac{h_{n_1,n_2}^{(2)}}{h_{n_1,n_2-1}^{(2)}},
    \end{eqnarray}
and
\begin{eqnarray} \label{formulacn-1n+1}
c_{n-1,n+1} = \frac{h_{n_1+1,n_2}^{(2)}}{h_{n_1,n_2-1}^{(2)}}.
\end{eqnarray}

Now we plug all formulas (\ref{relationP}), (\ref{relationQ}),
(\ref{formulacn-2n}), (\ref{formulacn-1n}) and (\ref{formulacn-1n+1})
into (\ref{formulaKn2}). Straightforward calculations then lead
to (\ref{CDformula}).
\end{proof}

\section{Riemann-Hilbert problem}
We use the notation of Section 4.

The Christoffel-Darboux formula (\ref{CDformula}) can be expressed
in terms of the solution of a Riemann-Hilbert problem that was given
by Van Assche, Geronimo, and Kuijlaars \cite{VAGK} to characterize the multiple
orthogonal polynomials, and which generalizes the Riemann-Hilbert problem
for orthogonal polynomials due to Fokas, Its, and Kitaev \cite{FIK}.
The Rieman-Hilbert problem is to find
$Y : \mathbb C \setminus \mathbb R \to \mathbb C^{3\times 3}$
such that
\begin{itemize}
\item $Y$ is analytic on $\mathbb C \setminus \mathbb R$,
\item for $x \in \mathbb R$, we have
\begin{equation} \label{jumpY}
    Y_+(x)  = Y_-(x) \begin{pmatrix} 1 & w_1(x) & w_2(x) \\
    0 & 1 & 0 \\ 0 & 0 & 1 \end{pmatrix}
\end{equation}
where $Y_+(x)$ ($Y_-(x)$) denotes the limit of $Y(z)$ as $z \to x$ from
the upper (lower) half-plane,
\item as $z \to \infty$, we have
\begin{equation} \label{asympY}
    Y(z) = \left(I + O\left(\frac{1}{z}\right)\right)
    \begin{pmatrix} z^{n} & 0 & 0 \\
    0 & z^{-n_1} & 0 \\ 0 & 0 & z^{-n_2} \end{pmatrix}
\end{equation}
where $I$ denotes the $3\times 3$ identity matrix.
\end{itemize}
In \cite{VAGK} it was shown showed that there
is a unique solution
\begin{equation} \label{solutionY}
Y = \begin{pmatrix}
    P_{n_1,n_2} & C(P_{n_1,n_2} w_1) & C(P_{n_1,n_2} w_2) \\[10pt]
    c_1 P_{n_1-1,n_2} & c_1 C(P_{n_1-1,n_2}w_1) & c_1 C(P_{n_1-1,n_2} w_2) \\[10pt]
    c_2 P_{n_1,n_2-1} & c_2 C(P_{n_1,n_2-1}w_1) & c_2 C(P_{n_1,n_2-1} w_2)
    \end{pmatrix}
\end{equation}
with constants
\begin{equation} \label{c1c2}
c_1 = -2\pi i \left( h_{n_1-1,n_2}^{(1)} \right)^{-1}, \quad
\textrm{and} \quad
c_2 = -2\pi i \left(h_{n_1,n_2-1}^{(2)}\right)^{-1},
\end{equation}
 and where
$Cf$ denotes the Cauchy transform of $f$, i.e.,
\[ Cf(z) = \frac{1}{2\pi i} \int_{\mathbb R} \frac{f(s)}{s-z} ds. \]

The multiple orthogonal polynomials of type I $A_{n_1,n_2}$, $B_{n_1,n_2}$
have a Riemann-Hilbert characterization as well. We seek
$X : \mathbb C \setminus \mathbb R \to \mathbb C^{3\times 3}$
such that
\begin{itemize}
\item $X$ is analytic on $\mathbb C \setminus \mathbb R$,
\item for $x \in \mathbb R$, we have
\begin{equation} \label{jumpX}
    X_+(x) = X_-(x) \begin{pmatrix} 1 & 0 & 0 \\
   -w_1(x) & 1 & 0 \\
   -w_2(x) & 0 & 1 \end{pmatrix}
\end{equation}
\item as $z \to \infty$, we have
\begin{equation} \label{asympX}
    X(z) = \left(I + O\left(\frac{1}{z}\right)\right)
    \begin{pmatrix} z^{-n} & 0 & 0 \\
    0 & z^{n_1} & 0 \\ 0 & 0 & z^{n_2} \end{pmatrix}.
\end{equation}
\end{itemize}

The solution to this Riemann-Hilbert problem \cite{VAGK} is
\begin{equation} \label{solutionX}
X = \begin{pmatrix}
    -2 \pi i C(A_{n_1,n_2} w_1 + B_{n_1,n_2} w_2) &
    2\pi i A_{n_1,n_2} & 2\pi i B_{n_1,n_2} \\[10pt]
    -k_1 C(A_{n_1+1,n_2}w_1 + B_{n_1+1,n_2} w_2)  &
    k_1 A_{n_1+1,n_2} & k_1 B_{n_1+1,n_2} \\[10pt]
    -k_2 C(A_{n_1,n_2+1}w_2 + B_{n_1,n_2+1} w_2) &
    k_2 A_{n_1,n_2+1} & k_2 B_{n_1,n_2+1} \end{pmatrix}
\end{equation}
where
\begin{equation} \label{k1}
k_1 = \frac{1}{\textrm{leading coefficient of }A_{n_1+1,n_2}} = h_{n_1,n_2}^{(1)},
\end{equation}
and
\begin{equation} \label{k2}
k_2 = \frac{1}{\textrm{leading coefficient of }B_{n_1,n_2+1}} = h_{n_1,n_2}^{(2)}.
\end{equation}

It is easy to see that
\begin{equation} \label{relationXY}
    X = Y^{-t} \qquad \textrm{(inverse transpose)}
\end{equation}
see also \cite{VAGK}.

Now we form the product $Y^{-1}(y) Y(x) = X^t(y) Y(x)$ and we compute
the $21$-entry using (\ref{solutionY}), (\ref{c1c2}), (\ref{solutionX}), (\ref{k1}),
(\ref{k2}),
\begin{eqnarray*}
\lefteqn{    [Y^{-1}(y) Y(x)]_{21} =
    \begin{pmatrix} 2 \pi i A_{n_1,n_2}(y) & k_1 A_{n_1+1,n_2}(y) & k_2 A_{n_1,n_2+1}(y) \end{pmatrix}
    \begin{pmatrix} P_{n_1,n_2}(x) \\ c_1 P_{n_1-1,n_2}(x) \\ c_2 P_{n_1,n_2-1}(x) \end{pmatrix} }\\
    & & =  2\pi i \left(P_{n_1,n_2}(x) A_{n_1,n_2}(y)
        - \frac{h_{n_1,n_2}^{(1)}}{h_{n_1-1,n_2}^{(1)}} P_{n_1-1,n_2}(x) A_{n_1+1,n_2}(y)
        - \frac{h_{n_1,n_2}^{(2)}}{h_{n_1,n_2-1}^{(2)}} P_{n_1,n_2-1}(x) A_{n_1,n_2+1}(y)
        \right).
\end{eqnarray*}
We get a similar expression for the $31$-entry $[Y^{-1}(y) Y(x)]_{31}$, but with the
$B$-polynomials instead of the $A$-polynomials.
Then it follows that we can rewrite the Christoffel-Darboux formula (\ref{CDformula})
as
\begin{equation}  \label{CDformula2}
    K_n(x,y) = e^{-\frac{1}{2} (V(x) + V(y))}
    \frac{e^{a_1y} [Y^{-1}(y) Y(x)]_{21}
        + e^{a_2y} [Y^{-1}(y) Y(x)]_{31}
        }{2\pi i(x-y)}
\end{equation}
which is a compact form for the kernel in terms of the solution of the
Riemann-Hilbert problem.
We expect that the Riemann-Hilbert problem for $Y$ is tractable to asymptotic analysis
using the methods of \cite{BI1,BI2} or \cite{Deift,DKMVZ1,DKMVZ2}.

\section*{Acknowledgments} The first author was supported in part by
NSF Grant DMS-9970625.
The second author was supported by FWO research grant G.0176.02, by
INTAS project 00-272, and by the Ministry of Science and Technology
(MCYT) of Spain, project code BFM2001-3878-C02-02.
He is grateful to Peter Dragnev (IUPUI-Fort Wayne)
for making a visit to Indiana possible.

\bigskip \bigskip

\noindent
P.M. Bleher: Department of Mathematical Sciences,
Indiana University-Purdue University Indianapolis,
402 N.~Blackford Street, Indianapolis, IN 46202, U.S.A.

\noindent
E-mail address: {\tt bleher@math.iupui.edu}
\bigskip

\noindent
A.B.J. Kuijlaars: Department of Mathematics, Katholieke
Universiteit Leuven, Celestijnenlaan 200 B, 3001 Leuven, Belgium

\noindent
E-mail address: {\tt arno@wis.kuleuven.ac.be}

\end{document}